\documentclass[12pt]{iopart}
\usepackage{iopams}  
\usepackage{xcolor} 
\usepackage{graphicx}
\usepackage{hyperref} 
\usepackage{dirtytalk}
\expandafter\let\csname equation*\endcsname\relax
\expandafter\let\csname endequation*\endcsname\relax
\usepackage{amsmath}

\begin{document}
\UseRawInputEncoding
\newcommand{\WSe}{WSe\textsubscript{2}}
\newcommand{\MX}{MX\textsubscript{2}}

\title[\WSe~on Au]{Electronic structure of the interface between Au and \WSe}

\author{Laxman Nagireddy}
\address{Department of Physics, University of Warwick, Coventry, CV4 7AL, UK}
\address{CY Cergy Paris Université, CEA, LIDYL, 91191 Gif-sur-Yvette, France}
\address{Université Paris-Saclay, CEA, LIDYL, 91191, Gif-sur-Yvette, France}

\author{Samuel J. Magorrian}
\address{Department of Physics, University of Warwick, Coventry, CV4 7AL, UK}

\author{Matthew D. Watson}
\address{Diamond Light Source Ltd, Harwell Science and Innovation Campus, Didcot, OX11 0DE, UK}

\author{Yogal Prasad Ghimirey}
\address{Diamond Light Source Ltd, Harwell Science and Innovation Campus, Didcot, OX11 0DE, UK}
\address{Department of Physics, University of Warwick, Coventry, CV4 7AL, UK}

\author{Marc Walker}
\address{Photoemission Research Technology Platform, Department of Physics, University of Warwick, Coventry, CV4 7AL, UK}

\author{Cephise Cacho}
\address{Diamond Light Source Ltd, Harwell Science and Innovation Campus, Didcot, OX11 0DE, UK}

\author{Neil R. Wilson and Nicholas D. M. Hine}
\address{Department of Physics, University of Warwick, Coventry, CV4 7AL, UK}
\ead{Neil.Wilson@warwick.ac.uk, N.D.M.Hine@warwick.ac.uk}

\newpage

\begin{abstract}

Understanding the interface between metals and two-dimensional materials is critical for their application in electronics and for the development of metal-mediated exfoliation of large area monolayers. Studying the intricate interactions at the interface requires model systems that enable control of the roughness, purity, and crystallinity of the metal surface. Here, we investigate the layer-dependent electronic structure of \WSe~on template-stripped gold substrates fabricated using both silicon and mica templates, giving crystallographically disordered and Au(111) ordered surfaces, respectively, and contrast these findings with \textit{ab initio} predictions. We observe strong hybridization around the Brillouin zone centre at $\overline{\Gamma}$, indicating a covalent admixture in the gold-\WSe~interaction, and band shifts that suggest charge rearrangement at the Au(111) / \WSe~interface. Core-level spectroscopy shows a single chemical environment for the interfacial \WSe~layer on the template-stripped gold, distinct from the subsequent layers. These results reveal a mixture of van der Waals and covalent interactions, best described as a covalent-like quasi-bonding with intermediate interaction strength.
\end{abstract}


\vspace{2pc}
\noindent{{\it \today}}

\section{Introduction}
There has been a rapid development in electronics and optoelectronics of two-dimensional (2D) semiconductors \cite{Schaibley2016, Furchi2014PhotovoltaicHeterojunction}, with the semiconducting transition metal dichalcogenide family (\MX, with M = Mo or W and X = S or Se) leading the way \cite{Wang2012ElectronicsDichalcogenides.,Manzeli20172DDichalcogenides}. Developing electronic devices requires integration with other materials, such as metals for electrical contacts. Research into metal - 2D semiconductor junctions has dramatically decreased contact resistance \cite{Wang2021MakingDichalcogenides,Zheng2021OhmicMaterials}, showing the importance of minimising chemical disorder \cite{Liu2018ApproachingJunctions,Wang2019VanSemiconductors}, controlling Fermi-level pinning \cite{Liu2022FermiProspects}, suppressing metal-induced gap states \cite{Shen2021UltralowSemiconductors}, minimising the Schottky barrier height \cite{Liu2018ApproachingJunctions} which depends on the energy level alignments at the interface, and reducing the van der Waals' gap \cite{Li2023ApproachingContacts}. 

Understanding the metal - 2D semiconductor interface is therefore of critical importance to the development of the field and requires the study of well-defined systems. As such, there is great interest in the interface between Au and \MX~\cite{Velicky2020TheSubstrates,Pirker2024WhenMetals}. Recent results have also shown that the dependence of the electronic interaction on the Au crystallography must also be taken into account \cite{Zhu2022VisualizingSurfaces}, resulting in nanoscale heterogeneity for typical polycrystalline Au surfaces \cite{Boehm2023EngineeringInterfaces}.

The strong interaction between Au and \MX~has led to another area of research, the fabrication of large-area monolayers of \MX~by gold-mediated exfoliation \cite{Magda2015ExfoliationLayers, Desai2016Gold-MediatedMonolayers,Velicky2018MechanismMonolayers,Huang2020UniversalCrystals,Heyl2020ThermallyTransfer,Liu2020DisassemblingLattices,Li2021DryArrays,Gramling2019SpatiallySources,Heyl2023OnlyTape}. Following the pioneering work of Magda et al. \cite{Magda2015ExfoliationLayers}, Huang et al. found that exfoliation required an intermediate interaction strength between the 2D material and the metal, covalent-like quasi-bonding (CLQB), to provide strong adhesion without fundamentally altering the properties of the 2D layer \cite{Huang2020UniversalCrystals}. With this, they demonstrated the exfoliation of 40 different two-dimensional materials (2DMs) \cite{Huang2020UniversalCrystals}. Gold-mediated exfoliation has been further developed to create large-area heterostructure arrays \cite{Li2021DryArrays} and 2D artificial lattices \cite{Liu2020DisassemblingLattices}. 

Maximizing direct atomic contact between metal and 2DM is essential for effective exfoliation. This leads to two critical requirements: the gold surface must be flat and clean. Gold is rapidly contaminated on exposure to air, by hydrocarbons and other airborne contaminants. Hence, the gold surface should be used within a few minutes of exposure to the atmosphere \cite{Velicky2018MechanismMonolayers}. A simple approach to prepare such ultra-flat gold is through the use of an ultra-flat template substrate \cite{Vogel2012AsProcedures}. Template-stripped gold (TSG) was developed to provide flat metal surfaces to study molecular assembly and surface forces, for example, by surface science techniques such as scanning probe microscopy \cite{Hegner1993UltralargeMicroscopy,Vogel2012AsProcedures}. Similarly, exfoliation on TSG provides a convenient system for studying the \MX~- Au interface \cite{Pollmann2021Large-AreaInterface,Pushkarna2023Twist-Angle-DependentAu111}.

In this work, we employ angle-resolved photoemission spectroscopy with micrometre-scale spatial resolution (\textmu ARPES) to examine the electronic structure at the interface of \WSe~and template-stripped gold, and contrast these findings with \textit{ab initio} predictions. Both experimental and theoretical spectra indicate strong hybridization near the center of the Brillouin zone at $\overline{\Gamma}$, as well as shifts in the \WSe~bands on TSG. However, there is no evidence of occupation of the \WSe~conduction band or the presence of any \WSe~states at the Fermi energy.

\section{Results}

\begin{figure}[t]   
\centering
        \includegraphics[width=\linewidth]{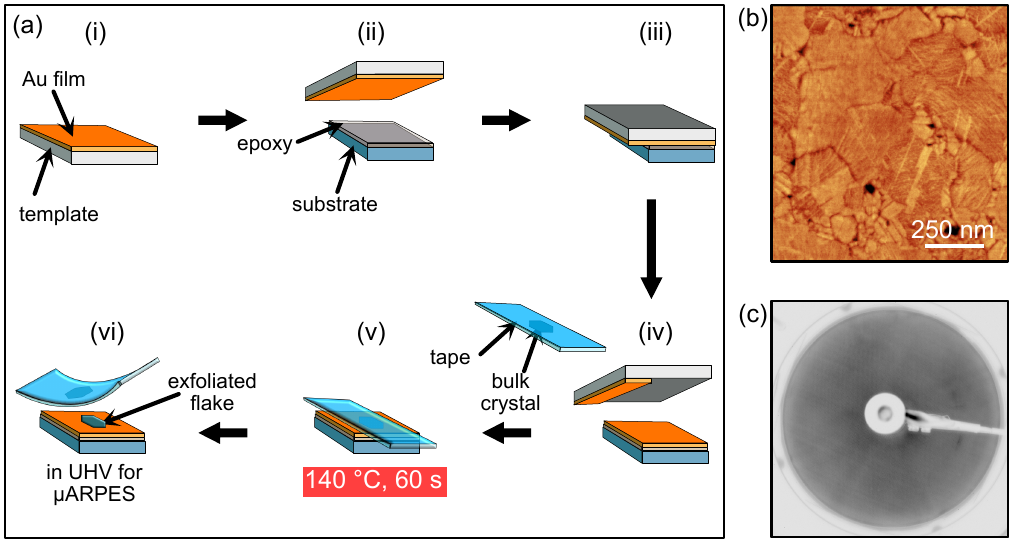}
    \caption{\textbf{Large-area exfoliation using template stripped gold}. (a) Schematic of the template stripped gold technique for exfoliation of 2D materials. (b) Atomic force microscopy height image of Si-TSG, full height scale 5 nm. (c) Low energy electron diffraction of Si-TSG, taken with an electron beam energy of 150 eV.}
    \label{fig:TSG}
\end{figure}

The sample fabrication process is schematically described in Figure \ref{fig:TSG}a), and the Methods section, and follows well-established protocols for the preparation of template-stripped gold \cite{Hegner1993UltralargeMicroscopy, Vogel2012AsProcedures}, using silicon (Si-TSG) and mica (mica-TSG) as templates. First, a thin gold film (typically around 100 nm) is deposited on the template substrate, Figure \ref{fig:TSG}a)(i), and then glued to a silicon carrier substrate, Figure 1a)(ii) and (iii). Immediately prior to use, the template is removed to reveal the template-stripped gold substrate. Atomic force microscopy (AFM) shows that Si-TSG gives a uniform, flat, polycrystalline, gold surface with a typical surface roughness of 4.0 $\pm$ 0.6~\AA, Figure \ref{fig:TSG}b). Low energy electron diffraction (LEED) patterns from Si-TSG show no diffraction spots (Figure \ref{fig:TSG} c), and hence there is no evidence for a preferred crystallographic surface orientation or long-range order.  

Exfoliating onto TSG gives large thin, often monolayer, flakes of many layered crystals. Efficient exfoliation requires that the TSG is used as soon as possible after the template is removed \cite{Velicky2018MechanismMonolayers}, Figure 1a)(iv). Furthermore, as reported by Heyl et al. \cite{Heyl2020ThermallyTransfer}, we found that thermal activation by heating to 140$^\circ$C was necessary to activate the interface between the layered crystal and gold before exfoliation, Figure 1a)(v). For \textmu ARPES, the final step of the exfoliation, Figure 1a)(vi), was carried out under ultra-high vacuum (UHV) to retain a clean flake surface after exfoliation. 

\begin{figure}[t]   
\centering
        \includegraphics[width=\linewidth]{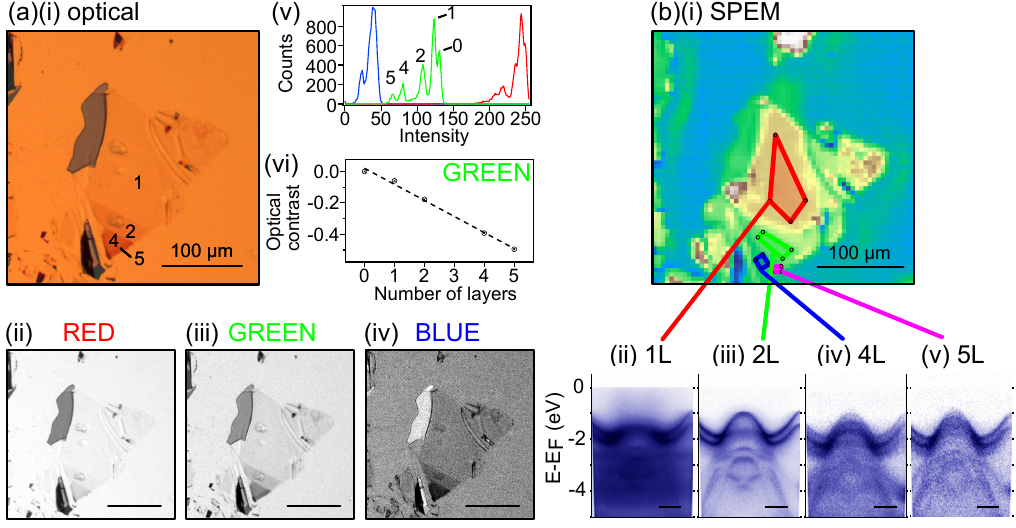}
    \caption{\textbf{Microscopy of \WSe~layers on Si-TSG}. (a)(i) Optical microscopy image of a \WSe~flake on Si-TSG, regions of different thickness are marked with the number of layers; this image is split into its red, green and blue components in (a)(ii), (iii) and (iv) respectively (scale bars are 100 \textmu m). (v) Histograms of the intensity counts from a region of this image, the blue line is from the blue image, green from green, and red from the red image. The peaks in the green line are labelled with the number of layers they correspond to. (vi) Optical contrast in the green channel plotted as a function of the number of layers, the dashed line is a linear fit to the data. (b)(i) Scanning photoemission microscopy image of the same region as in (a). Energy-momentum spectra averaged over the marked colored regions are shown in (ii)-(v) corresponding to 1L, 2L, 4L and 5L thick \WSe; the scale bars are 0.5 $\mathrm{\text{\AA}^{-1}}$.}
    \label{fig:opticalandspem}
\end{figure}

For the \WSe~crystals studied here, this approach reproducibly gives large flakes, typically on the order of hundreds of micrometers 
across \cite{Velicky2018MechanismMonolayers,Huang2020UniversalCrystals}, as shown in Figure \ref{fig:opticalandspem}. The largest regions of the \WSe~flakes are usually monolayers (1L), but there are also frequently two-layer (2L) and thicker regions, which facilitate the study of layer-dependent effects. In optical microscope images, such as Figure \ref{fig:opticalandspem}a), there is clear contrast between the \WSe~flakes and the gold surface. The optical contrast is greatest in the green component of the image: at almost 10\% for a monolayer, the optical contrast is comparable to that of \MX~monolayers on silicon oxide of optimised thickness \cite{Blake2007MakingVisible, Rubio-Bollinger2015EnhancedEasier, Bing2018OpticalMaterials}, and increases linearly with the number of layers of \WSe, giving a quick and convenient method for characterising the exfoliation. However, for \textmu ARPES, it was not possible to obtain high-resolution optical microscopy images of the flakes before acquiring spectra. Instead, using the \textmu ARPES optics, scanning photoemission microscopy (SPEM) was used to identify and map suitable flakes. SPEM can be used to unambiguously identify 1L, and 2L regions by their characteristic valence band dispersion around $\overline{\Gamma}$ \cite{Nguyen2019}: the quantum well nature of the 2D flakes means that the number of states in the upper valence band at $\overline{\Gamma}$ is equal to the number of layers, as shown in Figure \ref{fig:opticalandspem}b)(ii)-(v).

\begin{figure}[!ht]   
\centering
        \includegraphics[width=\linewidth]{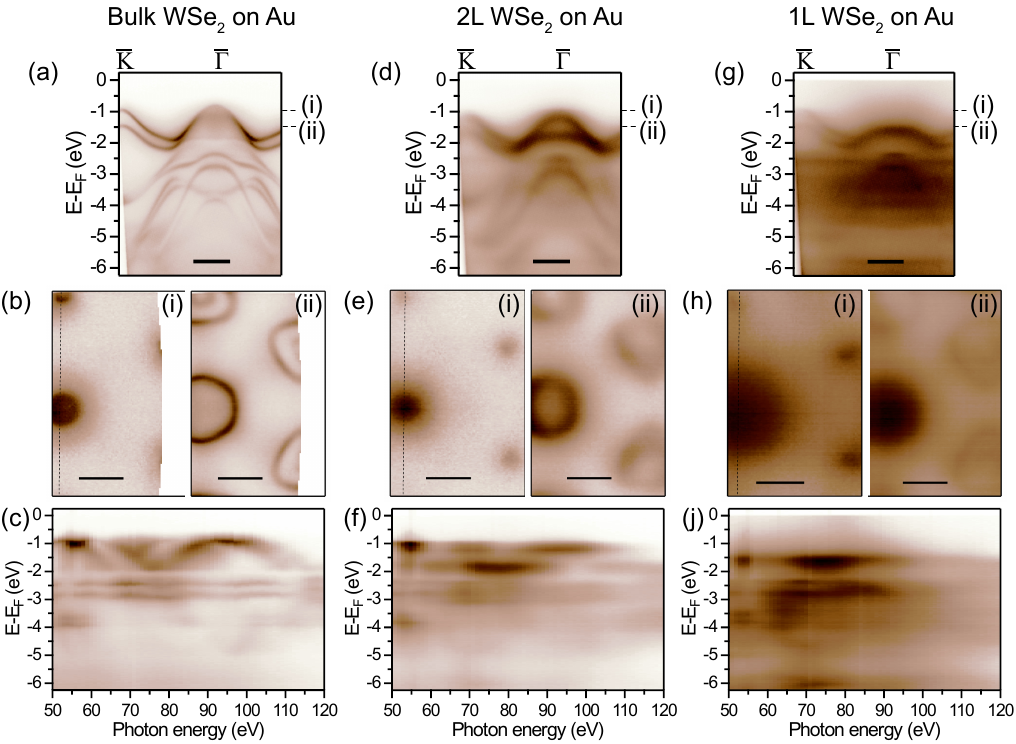}
    \caption{\textbf{\textmu ARPES of \WSe~on Si-TSG}. (a)/(d)/(g) Energy-momenta and (b)/(e)/(h) constant energy slices acquired on bulk / 2L / and 1L \WSe~on silicon TSG. The energy momenta slices are in the $\overline{\Gamma}$-$\overline{\mathrm{K}}$ direction, as marked by the dashed lines on the constant energy slices. The constant energy slices are averaged over 0.1 eV at (i) $\text{E}-\text{E}_\text{F}=-1.0$ eV and (ii) $\text{E}-\text{E}_\text{F}=-1.5$ eV as marked by the dashed lines in the energy-momentum slices. The scale bars are $\mathrm{\text{\AA}^{-1}}$. (c)/(f)/(j) Corresponding photon energy dependence of the energy distribution curves at $\overline{\Gamma}$ for bulk / 2L / and 1L \WSe~on silicon TSG.}
    \label{fig:SiTSG}
\end{figure}

ARPES spectra acquired from 1L, 2L and bulk regions of the same flake of \WSe~on Si-TSG are shown in Figure \ref{fig:SiTSG}, with constant energy maps in Figure \ref{fig:SiTSG}a), energy momentum (E-k) slices in b) and photon energy dependent spectra in c). The constant energy maps, at a binding energy of E-E$_{\mathrm{F}}= 1.5$ eV, are as expected for \WSe~\cite{Riley2015}, with intensity around the zone corners at $\overline{\mathrm{K}}$ and at the zone centre $\overline{\Gamma}$. 

For 2L and bulk \WSe, the E-k slices shown in Figure \ref{fig:SiTSG}b), in the $\overline{\Gamma} \rightarrow \overline{\mathrm{K}}$ direction, are roughly consistent with previous reports for \WSe~on insulating substrates such as hBN \cite{Wilson2017,Nguyen2019}. The upper valence band is split at $\overline{\mathrm{K}}$; the measured splittings, 0.45 $\pm$ 0.05 eV for 2L and 0.50 $\pm$ 0.05 eV for bulk, are consistent with theoretical predictions \cite{Kormanyos2015KSemiconductors} and previous experimental reports, such as 0.50 $\pm$ 0.01 eV in ref \cite{Nguyen2019}. The similarity in the upper valence band at $\overline{\mathrm{K}}$ in 2L and bulk \WSe~reflects the orbital character of the states around $\overline{\mathrm{K}}$. They are composed primarily of in-plane W 5\textit{d}$_{xy}$ and 5\textit{d}$_{x^2-y^2}$ orbitals \cite{Kormanyos2015KSemiconductors} and therefore change very little as the number of layers increases. In contrast, comparing the 2L and bulk spectra around the valence band edge at $\overline{\Gamma}$, clear differences are apparent. Here, the bands are primarily formed from Se 4\textit{p}$_z$ and W 5\textit{d}$_{z^2}$ orbitals and correspondingly have strong out-of-plane components that result in significant changes with increasing number of layers. In general, the edge of the valence band at $\overline{\Gamma}$ is pushed up to lower binding energies as the number of layers increases. In bulk material, this corresponds to a k$_z$ dispersion of the upper valence band states at $\overline{\Gamma}$, as can be seen in the photon energy-dependent spectra in Figure \ref{fig:SiTSG}c). For 2L, a similar pattern of intensity changes with photon energy is observed, but with quantum confinement effects quantizing the levels such that the bands are clearly 2D, not dispersing with the photon energy \cite{Salazar2025Two-dimensionalStructure}. Such changes in the photon energy dependence from bulk to 2D have been explained by Strocov \cite{Strocov2018PhotoemissionStates} in terms of Fourier analysis of the Bloch-wave components. For 2L, the valence band edges at  $\overline{\Gamma}$ and $\overline{\mathrm{K}}$ are at approximately the same binding energy, while for the bulk spectra the valence band maximum (VBM) is at $\overline{\Gamma}$: again, this is consistent with previous reports for \WSe~on an insulating substrate \cite{Nguyen2019}. 

For the spectra from 2L \WSe~on Si-TSG, we observe two differences compared to 2L \WSe~on hBN. Firstly, there is a weak background intensity that can be ascribed to photoemission from the gold substrate, attenuated through the \WSe~layers. This gives enough signal at the Fermi energy, E$_\mathrm{F}$, to allow E$_\mathrm{F}$ to be determined directly and accurately by fitting a Fermi function to the Fermi edge. Secondly, the bands are rigidly shifted in energy, with the valence band edge at $\overline{\mathrm{K}}$ moving from -0.75 $\pm$ 0.01 eV on hBN to -1.19 $\pm$ 0.05 eV on Si-TSG. Note that the band gap of 2L \WSe~on hBN has been measured to be 1.51 $\pm$ 0.03 eV \cite{Nguyen2019}, much larger than the valence band offset measured here. Although some renormalisation of the band gap may be expected due to the different dielectric environment \cite{Raja2017CoulombMaterials, Waldecker2019RigidScreening,Chaves2020BandgapMaterials}, with a gold substrate compared to hBN, the conduction band edge will still be well above E$_\mathrm{F}$. 
Supporting this, the photoemission intensity is uniform at E$_\mathrm{F}$ with no additional intensity around $\overline{\mathrm{Q}}$ or $\overline{\mathrm{K}}$ and hence no evidence for occupation of the conduction band of the \WSe.

The ARPES spectra of 1L \WSe~in Si-TSG, Figure \ref{fig:SiTSG}g), show similar changes compared to 1L \WSe~on hBN. Although the upper valence band is still apparent, spin split at $\overline{\mathrm{K}}$ and merging to a single spin degenerate band at $\overline{\Gamma}$, in most parts of the spectra there is strong photoemission that is not present for \WSe~on hBN. Only in the region E-E$_\mathrm{F}$ = -2 eV to -1 eV is the photoemission primarily from the \WSe, corresponding to the upper valence band of the \WSe, with the VBM at $\overline{\mathrm{K}}$ at -1.08 $\pm$ 0.05 eV, significantly higher binding energy than the -0.80 $\pm$ 0.01 eV previously measured for 1L \WSe~on hBN \cite{Nguyen2019}. Increased photoemission intensity across a broad range of binding energies is also apparent in the photon-energy dependent spectra shown in Figure \ref{fig:SiTSG}j). However, the polycrystalline disordered gold surface of Si-TSG complicates identifying specific spectral features that may arise from interactions between gold and \WSe. 

\begin{figure}[t]
    \centering
    \includegraphics[width=0.7\textwidth]{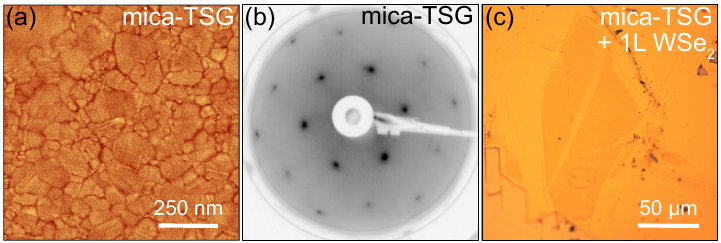}
   \caption{\label{fig:mica-TSG-surface}  mica-TSG surface characterisation. (a) Atomic force microscopy of mica-TSG, full height scale 5 nm. (b) Low energy electron diffraction of mica-TSG, taken with an electron beam energy of 150 eV. (c) Optical image of 1L \WSe~on mica-TSG. }
\end{figure}

For a better defined surface, we use mica as a template to give a flat ordered Au(111) surface. AFM shows that mica-TSG has a typical surface roughness of 2.6 $\pm$ 0.3~\AA, Figure \ref{fig:mica-TSG-surface} a). LEED measurements reveal the surface order, with a diffraction pattern consistent with unreconstructed Au(111), Figure \ref{fig:mica-TSG-surface}b). The pattern is acquired using an electron beam that is millimeters across; hence, the single set of diffraction spots shows not only atomic ordering at the surface but also long-range order in the Au(111) orientation due to epitaxy with the mica. Therefore, Au(111) on mica-TSG gives an interesting counterpart to the disordered Si-TSG surface.

As for Si-TSG, \WSe~can be exfoliated on to mica-TSG to give monolayers of large area, Figure \ref{fig:mica-TSG-surface} c). 
We note, however, a complication for the use of mica-TSG: removing the mica template from the substrate can leave thin layers of mica partially covering the surface \cite{Hegner1993UltralargeMicroscopy}. SPEM mapping allowed monolayer flakes on mica-TSG to be identified, avoiding the mica-covered regions. 

\begin{figure}[t]
    \centering
    \includegraphics[width=0.8\textwidth]{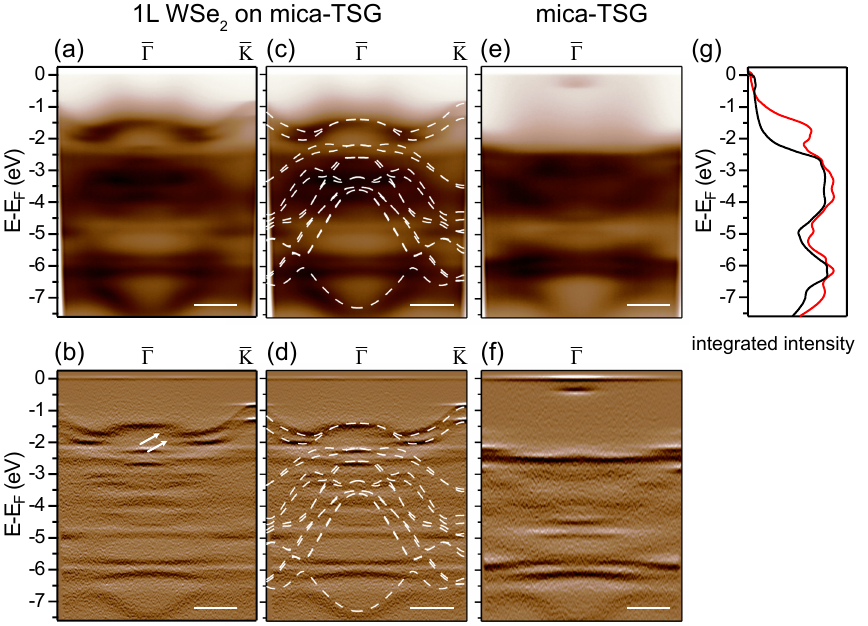}
   \caption{\label{fig:mica-ARPES}  \textbf{\textmu ARPES of \WSe~on mica-TSG}. Energy momenta slices of 1L \WSe~on mica-TSG,(a) and (c), and bare gold, (e), at 60 eV along the \textbf{$\overline{\Gamma}\rightarrow\overline{\mathrm{K}}$} high-symmetry direction, with the corresponding second derivative plots shown in (b), (d) and (e). The arrows in (b) indicate gaps in the dispersion. DFT calculated band dispersions are overlaid as white dashed lines in (c) and (d). The scale bars are $\mathrm{\text{\AA}^{-1}}$. Energy distribution curves, summed over the full $k$-range of the images in (a) and (e), are shown in (g); the red line is for 1L \WSe~on mica-TSG and the black for the bare gold surface.}
\end{figure}

Energy-momentum spectra from 1L \WSe~on mica-TSG, and exposed gold on mica-TSG, are compared in Figure \ref{fig:mica-ARPES}. Corresponding second derivative plots (numerical second derivative with respect to energy) are also shown to emphasize the band dispersions. In the 1L \WSe~E-k spectrum, the upper valence band of \WSe~is apparent between E-E$_\mathrm{F}$= -2 eV and -1 eV with clear spin-splitting of the bands at $\overline{\mathrm{K}}$, but the intensity in the rest of the spectrum is dominated by photoemission from the Au surface. The electronic structure of isolated 1L \WSe,  as calculated by density functional theory (DFT, data from \cite{Nguyen2019}), is overlaid on the \WSe~on mica-TSG spectrum and second derivative plot in Figures \ref{fig:mica-ARPES} c) and d). The DFT bands are known to agree well with the measured ARPES intensity from 1L \WSe~on hBN \cite{Nguyen2019}. Comparison between the second derivative plots and DFT shows that most aspects of the spectra are consistent with the isolated layers, but also highlights changes in the upper valence band of \WSe~around $\overline{\Gamma}$ where interaction between the gold and \WSe~bands is apparent. The band is broadened near $\overline{\Gamma}$ compared to around $\overline{\mathrm{K}}$  and the second derivative plot suggests that it is not continuous, with gaps indicated by arrows, around where the steeply dispersing gold bands cross the upper valence band of the \WSe~suggesting the importance of hybridisation between these bands. This observation is consistent with previous reports studying 1L MoS$_2$, and 1L WS$_2$, grown on Au(111) \cite{Miwa2015ElectronicMoS2,Dendzik2015GrowthAu111, Dendzik2017Substrate-induced/math}, and for \WSe~exfoliated onto Ag(111) in UHV using the \say{Kiss} method \cite{GrubisicCabo2023InMaterials}. Interestingly, this implies high interfacial quality in the \WSe~on mica-TSG, despite it being prepared \textit{ex situ}, not in UHV.

The valence band edge of the 1L \WSe~on mica-TSG is at E-E$_{\mathrm{F}} = -0.96 \pm 0.05$ eV, shifted by a small but significant amount relative to the E-E$_{\mathrm{F}} = -0.80 \pm 0.01$ eV measured on hBN \cite{Nguyen2019}. As for 1L \WSe~on Si-TSG, we attribute this band realignment to the formation of surface dipoles induced by charge rearrangement around the interface \cite{Gong2014TheInterfaces}. As for 1L \WSe~on Si-TSG, there is no evidence for occupied \WSe~states at the Fermi energy and therefore no evidence for charge transfer. Indeed, there is no photoemission intensity that can be attributed to the \WSe~at binding energies less than the VBM, also indicating that there is no evidence for gap states.

Comparison of intensity integrated energy-dispersion curves (iiEDCs) from the gold surface and 1L \WSe~shows no rigid band shift for the gold states, Figure \ref{fig:mica-ARPES}g). However, in the E-k spectrum from mica-TSG, there is photoemission intensity around $\overline{\Gamma}$ at E$_\mathrm{F}$. This is evidence for the surface state of gold that is known to arise from the herringbone reconstruction on Au(111) \cite{Barth1990ScanningDefects, Nicolay2002Spin-orbitAg111}. We note that this was only observed on a few areas of the mica-TSG samples, never on Si-TSG, and that it corresponded to areas of higher photoemission intensity than the rest of the exposed gold surface. As previously mentioned, cleaving the mica-TSG usually leaves some areas of the surface covered with thin layers of mica. We speculate that during the final exfoliation step in UHV, the Kapton tape can remove some of the thin mica, revealing a mica-TSG surface that has not been exposed to the glove box environment. 
No evidence of photoemission from a gold surface state was observed on mica-TSG with a \WSe~overlayer, consistent with previous reports that TMDs stabilize the unreconstructed surface on Au(111) \cite{Silva2022SpatialAu111, Grnborg2015SynthesisAu111}.
However, it appears that exposure of the gold surface to the glovebox environment is also sufficient to destabilize the reconstruction and surface state, as most of the exposed mica-TSG surface did not show defined photoemission intensity around $\overline{\Gamma}$ at E$_\mathrm{F}$. Hence, the absence here of the gold surface state under 1L \WSe~could be due to contamination of the gold surface.

\begin{figure}[!ht]
    \centering
    \includegraphics[width=0.8\textwidth]{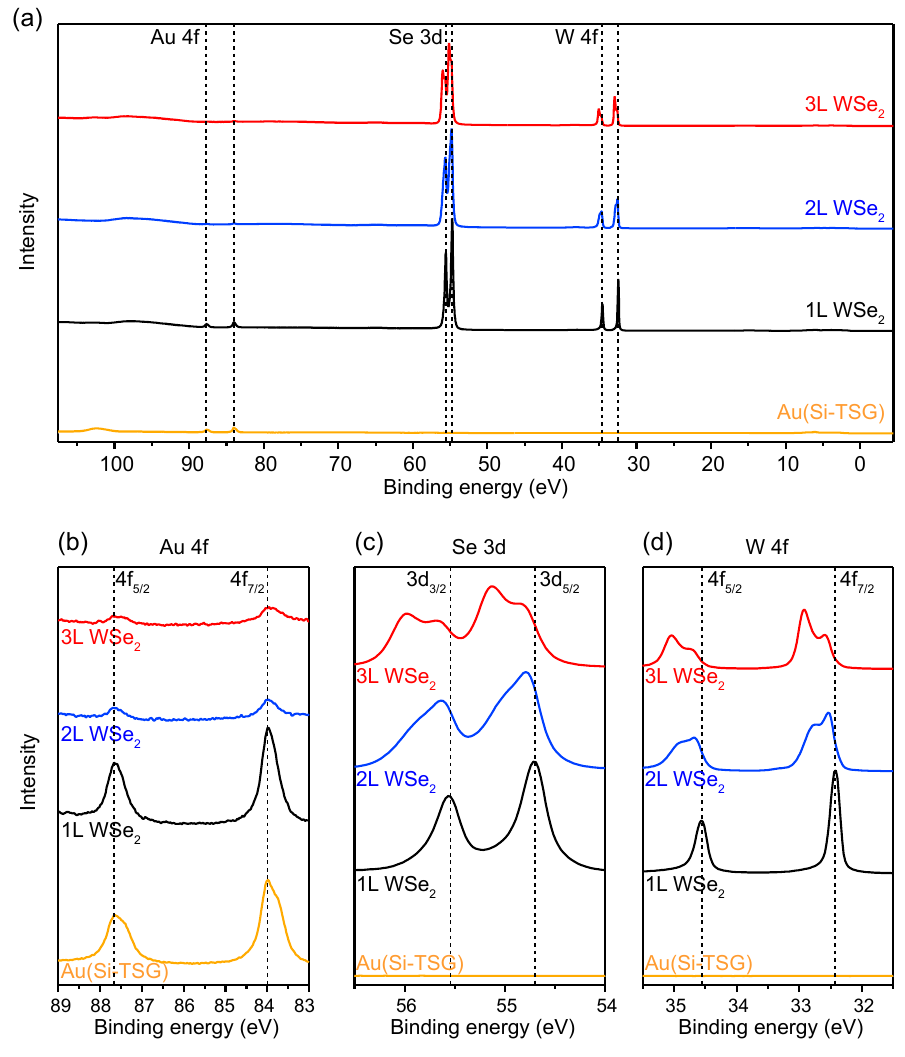}
   \caption{\label{fig:Si_TSG_corelevels} (a) Core level spectra recorded for 1L, 2L, 3L of \WSe~on Si-TSG and Au region, acquired at a photon energy of 140 eV. (b-d) Individual spectra of  Au 4\textit{f}, Se 3\textit{d} and W 4\textit{f} respectively.}
\end{figure}

Further insight can be gained from core-level photoemission spectra, as shown in Figure \ref{fig:Si_TSG_corelevels} for \WSe~on Si-TSG (similar conclusions can be drawn from data for \WSe~on mica-TSG, as shown in the supplementary data). The core level spectra are plotted versus binding energy, where the binding energy $\text{BE}=-\left(\text{E}-\text{E}_\text{F}\right)$. The survey scan, Figure \ref{fig:Si_TSG_corelevels} a), shows characteristic core levels from W 4\textit{f }(30 to 40 eV), Se 3\textit{d} (50 to 60 eV), and Au 4\textit{f} (80 to 90) eV. In some spectra, there is also a peak around 100 eV which we assign to Si 2\textit{p}.  

The corresponding region scans, shown in Figures \ref{fig:Si_TSG_corelevels} 
 b)-d), reveal further details about the interaction between the gold substrate and the \WSe~overlayers. For 1L \WSe~a single chemical environment is observed for W and Se atoms, with single pairs of spin-split W 4\textit{f} and Se 3d peaks. 
 However, for 2L \WSe, a second set of peaks is resolved with a higher binding energy in both the W 4\textit{f} and Se 3\textit{d} regions, indicating that the second layer is in a chemical environment different from the first. Consistent with this, the spectra from 3L \WSe~also show two sets of peaks, with a higher intensity in the higher binding energy peaks. The separation between pairs of peaks is 0.21 $\pm$ 0.02 eV on average, consistent across W 4\textit{f} and Se 3\textit{d}, and between 2L and 3L \WSe. (A summary of the peak energies is given in a table in the supplementary data.) The key conclusion is that there is a single chemical environment for the W and Se atoms in the monolayer on gold, which is distinct from the subsequent layers. The single chemical environment for the Se atoms in the monolayer is noteworthy: for MoS$_2$ on Au(111), a previous report of the XPS core-level analysis showed that the chalcogen atom at the gold interface, $S_{bottom}$, is in a different chemical environment from the upper chalcogen atoms $S_{top}$, indicating strong hybridization between the gold surface atoms and $S_{bottom}$ \cite{Bana2018EpitaxialMonolayers}.
 
 Au 4\textit{f} peaks are observed at the same energies 
 for bare gold, 1L, 2L and 3L \WSe, but with a reducing intensity for an increasing number of \WSe~layers, as expected due to attenuation. The similarity in the Au 4\textit{f} peak positions is consistent with the absence of peak shifts in the Au valence bands, as discussed earlier, Figure \ref{fig:mica-ARPES}g). Finally, an unexpected peak is observed at around 102 eV, most prominently on the bare gold surface. This was not studied in detail, but its intensity was found to vary at different positions even on the bare TSG surface. Some photoemission intensity around this energy was also found under the \WSe. We assign this peak to Si 2\textit{p}. We note that Si 2\textit{p} has a high cross section: therefore, although the peak is relatively intense, it does not signify a layer of Si, but rather a small amount of contamination. It is not clear where this contamination comes from: the gold surface was stripped from a silicon substrate, briefly exposed to the glovebox environment, and may have contacted the Kapton tape used for exfoliation, which has an adhesive containing silicon. Regardless of its source, the observation of this contaminant reinforces that TSG surfaces are not carefully prepared in UHV and hence always exhibit some disorder. However, it is interesting to compare these results with the model case of 1L \WSe~on Au(111).

\begin{figure}[!t]   
\centering
        \includegraphics[width=\linewidth]{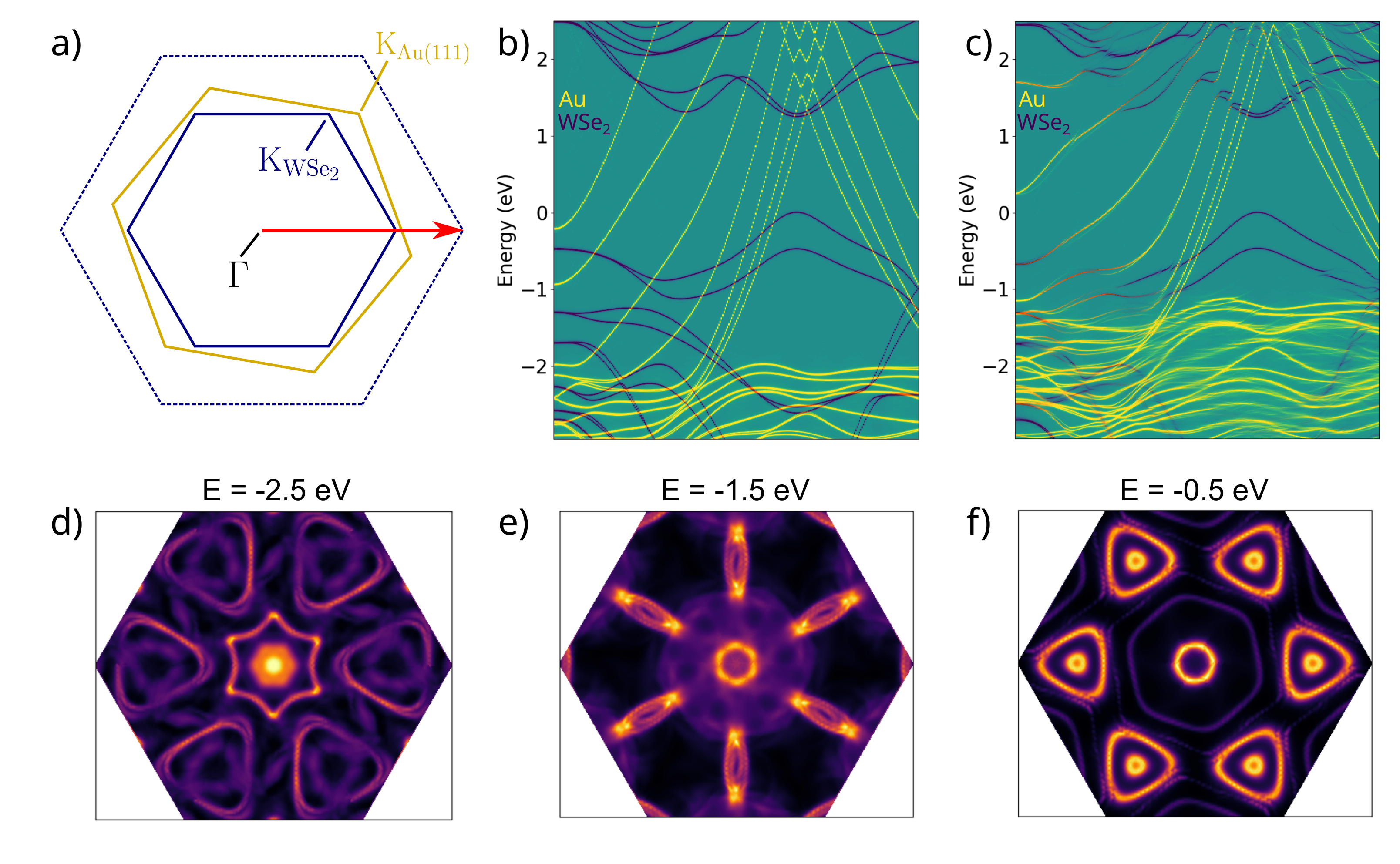}
    \caption{\textbf{DFT bandstructure for \WSe~on Au(111)}. (a) Schematic of 2D Brillouin zones of 1L \WSe~(blue) and the Au(111) surface (gold), with relative twist angle between them of 9.8$^{\circ}$. The red arrow indicates the path along which unfolded bands are plotted in (b) and (c) and the dashed hexagon gives the extent of constant energy maps in (d)-(f). (b) DFT bands for isolated monolayer \WSe~(blue lines) and the 6-layer Au(111) slab (yellow lines), E$_\mathrm{F}$ = -4.88 eV. (c) Bands for the \WSe~/ Au(111) heterostructure, unfolded and projected onto the  \WSe~primitive cell, E$_\mathrm{F}$ = -5.08 eV. The energy scales in (b) and (c) are relative to the valence band maximum of \WSe. Blue denotes states in \WSe~and yellow states in Au(111); states hybridised between the layers can be seen from their mixed color. (d)-(f) Constant energy maps for heterostructure bands unfolded and projected onto the \WSe~primitive cell, extending for 1.5 times the first Brillouin zone of \WSe, as indicated by the dashed hexagon in (a).}
    \label{fig:theory_AuwSe2}
\end{figure}

To elucidate the effect of hybridisation and charge transfer on the bandstructure, we perform large-scale density functional theory calculations of the interface between \WSe~and Au(111). Given that the band offsets in TMDs are notably sensitive to strain \cite{Johari2012, Desai2014, Schmidt2016, PhysRevB.98.115308}, it is essential to keep the strain minimal in the construction of interface models. Consequently, for a lattice-mismatched heterostructure, it is necessary to construct large surface supercells of each component to find a construction that is very nearly commensurate. Because the interface being modeled does not possess any defined relative alignment between the lattices, we are free to search over possible supercell constructions and twist angles to find a model with low strain. In this case, for the Au(111) surface with \WSe, we find that with a relative twist of $9.8^\circ$, all components of strain and shear can be kept under 0.15\%, which is low enough to not substantially change the \WSe~bandstructure. The electronic structure of the model comprising monolayer \WSe~and 6-layer Au(111) was simulated (see supplementary data for justification that 6-layers is sufficient), and interlayer distance optimisation was performed, after which separation of the topmost layer of Au from that of Se is 2.9 \AA. Calculations were also performed, for comparison, on the corresponding extracted geometries of the separated systems for the 1L \WSe~and the 6L Au(111) surface.

Spectral function unfolding was then performed on all three systems, to unfold the electronic bandstructure to the \WSe~primitive cell, as shown in Figure \ref{fig:theory_AuwSe2}. Although the results presented are for the specific twist angle of 9.8\textdegree, as depicted in the Brillouin zone schematic in Figure \ref{fig:theory_AuwSe2} a), the key results were independent of the twist angle. Comparing the spectral functions for the two isolated systems, Figure \ref{fig:theory_AuwSe2} b), with that of the combined system, Figure \ref{fig:theory_AuwSe2} c), emphasizes the effects of hybridization between Au and \WSe. The energy axis has been shifted to align the VBM of \WSe~at zero in both cases in the plots. On an absolute scale, the Fermi level is at energies of -4.88 eV and -5.08 eV for the isolated \WSe~and for the \WSe~/ Au(111) heterostructure respectively, with the \WSe~VBM at E($\overline{\mathrm{K}}$) = -5.11 eV and -5.53 eV. Consequently, the \WSe~VBM shifts by 0.22 eV from the isolated layer to the heterostructure, in agreement with our experimentally observed shift of 0.16 $\pm$ 0.05 eV between 1L \WSe~on mica-TSG compared to on hBN, which we assume to be electronically inert \cite{Magorrian2022BandHeterostructures}.

Because there are no Au(111) states at nearby energies, the states at $\overline{\mathrm{K}}$ at the VBM and CBM of \WSe~remain localised in the \WSe~layer, but there are significant interlayer interactions in the conduction band near the $\overline{\mathrm{Q}}$ point for the \WSe~/ Au(111) heterostructure. Most clearly, there is strong hybridization between the Au and \WSe~states in the upper valence band near $\overline{\mathrm{\Gamma}}$, with clear anticrossings and states of mixed character. These observations agree well with the experimental data, in which ARPES gives evidence for distortions of the electronic structure corresponding to hybridization between gold and \WSe~states around $\overline{\Gamma}$, while the rest of the electronic structure of the \WSe~is mostly retained but with a small rigid shift indicative of charge rearrangement at the interface \cite{Gong2014TheInterfaces}.

Figure \ref{fig:theory_AuwSe2}(d-f) show constant energy maps of the spectral function at slices across the extent of the blue dashed hexagon indicated in Figure \ref{fig:theory_AuwSe2} a). The general appearance agrees well with those in Figure \ref{fig:SiTSG}, and one can clearly observe the reduction from $D_6$ to $C_6$ symmetry, via breaking of mirror symmetries, following hybridization between overlapping Au and \WSe~bands.

In summary, both experimental and theoretical spectra give evidence for strong hybridization around the Brillouin zone centre at $\overline{\Gamma}$, indicating a covalent admixture in the gold-\WSe~interaction, and for charge rearrangement at the Au(111) /  \WSe~interface, but there is no evidence for occupation of the \WSe~conduction band or of any other \WSe~states at the Fermi energy.

\section{Discussion and conclusions}
Together, these findings provide fresh understanding of the interaction between gold and \WSe. We observe: a rigid shift of the \WSe~bands on TSG, compared to \WSe~on hBN; shifts in core levels solely within the \WSe~layer adjacent to the gold surface; a single chemical environment for the interfacial monolayer \WSe~on TSG, as evidenced by Se 3\textit{d} core level analysis, with a different chemical environment for the subsequent layers; evidence for hybridization between the gold and \WSe, particularly near $\overline{\Gamma}$ in the upper valence band of \WSe, observed in valence band spectra on the more ordered mica-TSG; as well as avoided crossings and mixed composition states in DFT predictions of the \WSe~-~gold interface's electronic structure. However, there is no experimental or theoretical evidence for filled states at or near the Fermi energy within the \WSe. 

This allows us to compare the strength of interaction with other previously reported systems of 2D materials on metals. As discussed by Pirker et al. \cite{Pirker2024WhenMetals}, the terminology used in the literature to qualify the strength of interaction at the 2DM-metal interface can be inconsistent. At the most basic level, for these systems, a weak interaction refers to a predominantly van der Waals (vdW) interaction that induces only a small change in the properties of the 2DM. In such interactions, there are no structural changes in the 2DM or metal surface, and the electronic structure of each is retained unchanged, superposed with only rigid shifts caused by charge transfer or interfacial charge rearrangement. Examples of weak metal-2DM interactions include graphene on Au(111) \cite{Wofford2012ExtraordinaryAu111} and graphene grown on copper \cite{Wilson2013, Li2009}. Conversely, a strong interaction induces a large change in the properties of the 2DM and is more covalent-like, i.e. electronic states are hybridised between the surface of the metal and the 2DM such that electrons are shared between them in covalent bonds. This results in a heavily distorted electronic structure, as found for graphene on Ru(0001) or Ni(111) \cite{Brugger2009ComparisonRu0001}.

From the results shown here, \WSe~on gold is an intermediate case. ARPES gives evidence for distortions of the electronic structure corresponding to hybridization between gold and \WSe~states, indicating a covalent admixture in the gold-\WSe~interaction, but only around $\overline{\Gamma}$. The rest of the \WSe~electronic structure is retained with a small rigid shift in binding energy that indicates some charge rearrangement at the interface. The core level spectra show that the interfacial layer of \WSe~is in a different chemical environment from subsequent layers, indicating a change in material properties and explaining why the TSG-mediated exfoliation of \WSe~can result in large monolayer areas \cite{Velicky2018MechanismMonolayers, Heyl2020ThermallyTransfer}. However, the core level spectra from 1L \WSe~on TSG do not show a difference in chemical shift between the interfacial chalcogenide atoms below the W atoms compared to the non-interfacial chalcogenide atoms above; such a shift was previously found for 1L MoS$_2$ epitaxially grown on Au(111) \cite{Grnborg2015SynthesisAu111}. The results presented here are therefore consistent with a mixture of vdW and covalent interactions for the \WSe-TSG interface, best described as a covalent-like quasi-bonding with intermediate interaction strength. 
 
 It is likely that such a mixture of vdW and covalent bonding is a generic and essential feature of metal-mediated exfoliation of 2DMs, although the extent of interfacial charge rearrangement or charge transfer can vary significantly. \WSe~on Au is a prototypical system; the same approach used here could be applied to a large number of 2D materials and a wide range of transition metal substrates (such as Pt, Pd, Cu, Ag, Au, Ir, Pb and so on) with varying work functions and interfacial bonding properties. This is, therefore, a tunable methodology for testing interfacial interactions in a format in which these fundamental properties can be directly probed.
 
 It is notable that the Au-\WSe~interface studied here was not formed in UHV, but through \textit{ex situ} exfoliation, and yet still showed clear evidence for a covalent admixture to the interaction, indicating a high interfacial quality. This suggests that control over the metal surface, as enabled by template-stripping, can be used to engineer consistent and desirable metal - 2D material interactions.

\section*{Methods}
\textbf{Sample fabrication:} magnetron sputtering was used to deposit gold onto a clean template surface. In this work, two templates were used, mica (AGG250-2 mica sheets from Agar Scientific) and silicon (phosphorous-doped silicon wafer pieces with native oxide only, from Inseto). The mica sheets were freshly cleaved with a razor blade immediately before being placed in the sputtering system. The silicon substrates were cleaned with acetone and isopropyl alcohol. For both templates, gold was deposited with a typical sputtering rate of 0.57 nm/s, to a total gold film thickness of about 100 nm. The sputtered gold surface was glued to a silicon carrier substrate using Opti-tec 5054-1 two-part epoxy (from Intertronics) and cured at 150$^\circ$C for 10 minutes. A fresh gold surface was exposed by cleaving away the template in an argon-filled glovebox with a razor blade. Within a minute of cleaving, flakes of bulk material on Kapton tape were pressed onto the fresh gold surface. The entire stack, including the Kapton tape, was placed on a hot plate at 140$^\circ$C for one minute. Finally, the Kapton tape was slowly peeled off, completing the exfoliation and leaving the TSG surface covered in thin flakes. The bulk crystals of \WSe~were purchased from HQ Graphene. For \textmu ARPES measurements, the Kapton tape was removed in UHV prior to transfer through UHV to the analysis chamber. \newline

\textbf{Atomic force microscopy:} images were acquired with a Bruker Icon operated in Peak Force tapping mode. \newline

\textbf{Low energy electron diffraction:} diffraction patterns were acquired with a SPECTALEED optic attached to a Multiprobe instrument, from Omicron Nanotechnology. \newline

\textbf{\textmu ARPES:} data were acquired at the nanoARPES branch of the i05 beamline at Diamond Light Source, UK. The beam was focused at a spot approximately 4 \textmu m in diameter using capillary mirror optics. Measurements were acquired with linear horizontal incident beam polarisation at a sample temperature of 30 K. Core-level spectra were acquired using the same optics, with a photon energy of 140 eV.\newline

\textbf{Density functional theory:} to establish the initial geometry of the monolayer \WSe~model, geometry optimisations were performed using the QuantumESPRESSO code \cite{Giannozzi2009, Giannozzi2017, Giannozzi2020}. We used the PBE exchange-correlation functional, pseudopotentials from the GBRV library of PAW potentials \cite{Blochl_1994}, a 12$\times$12 Monkhorst-Pack k-point grid \cite{Monkhorst_1976}, 80 Ry energy cutoff on the wavefunctions and 720 Ry energy cutoff on the density. For Au, following geometry optimisation in bulk, 3-layer, 6-layer and 9-layer Au(111) surface slab models were created using the ASE package \cite{Hjorth_Larsen_2017}, using the optimised bulk lattice constant. Slab models were geometry-optimised in QuantumEspresso, following which the electronic density of states (DoS) was calculated for the three slab thicknesses, and the 6-layer slab chosen as sufficient to converge the DoS (see Supplementary Figure 2). 

To create interface models of the \WSe~/ Au(111) interface, a search over possible constructions of near-coincident supercells with a relative twist in the range [0,30$^\circ$] was performed. The resulting supercells  for \WSe~and Au respectively had supercell matrices $\left[ \begin{smallmatrix} 9 & 0 \\ 0 & 9 \end{smallmatrix} \right]$ and $\left[\begin{smallmatrix} 11 & -2 \\ \phantom{-}9 & 11 \end{smallmatrix} \right]$  at an angle of 9.8$^\circ$. This near-perfect match required a uniform strain of just 0.13\% applied to the Au supercell, and near-vanishing shear components, to ensure an exact supercell match. 

Interface calculations were performed using the ONETEP Linear-Scaling Density Functional Theory package \cite{ONETEP2005,ONETEP2020}. The \WSe~lattice and the Au slab were kept in their relaxed geometries and interlayer distance optimisation was performed by translating the \WSe~model with respect to the 6-layer Au(111) model, with interlayer distances in the range 2.6-3.6~\AA, from which an optimal distance of 2.9~\AA~was chosen. ONETEP calculations were performed via simultaneous optimisation of a representation of the single-particle density matrix via a density kernel and a set of Non-orthogonal Generalised Wannier Functions (NGWFs) \cite{Skylaris2002}. NGWF radii were set to 13$a_0$, with 13 NGWFs/atom for W, 4 NGWFs/atom for Se and 9 NGWFs/atom for Au, using PAW potentials \cite{Hine2017} from the JTH library \cite{Jollet}, except for the case of Se where the potential was regenerated after minor adjustments to the JTH input for better smoothness of convergence within ONETEP (the resulting potential is attached). To ensure accurate treatment of interlayer interactions we used the optB88 van der Waals functional \cite{Klime_2009}. Ensemble DFT \cite{RuizSerrano2013} was used to enable a metallic treatment of the occupancies.  The underlying psinc grid on which the NGWFs are represented \cite{Mostofi2003} was set by a cutoff energy of 600~eV, with a density grid 4$\times$ finer. The Coulomb interaction was cutoff in a slab geometry \cite{Hine2011} at 42~\AA. 

At the optimised interlayer distance, a bandstructure calculation was performed using spectral function unfolding \cite{Popescu2012,Constantinescu2015,Graham2020GhostHeterostructures} to plot the Kohn-Sham eigenvalues unfolded to the underlying primitive cell of each layer. This was performed first along the reciprocal space path $\Gamma \rightarrow K_\mathrm{WSe_2} \rightarrow M_\mathrm{WSe_2}$, then on a dense grid of points in the irreducible wedge of the \WSe~Brillouin zone, in order to produce constant energy maps.

\section*{Data availability statement}
Data that support the plots in the manuscript are available from the corresponding authors on a reasonable request.

\section*{Acknowledgements}
We thank James Nunn, Karol Hricovini and Christine Richter for helpful discussions. We acknowledge support from EPSRC grant EP/T027207/1 and LN was supported by a EUTOPIA PhD Co-tutelle Programme. YPG acknowledges studentship support from EPSRC and Diamond Light Source. SJM and NDMH acknowledge funding from EPSRC grant number EP/V000136/1. MW acknowledges financial support from the EPSRC-funded Warwick Analytical Science Centre (EP/V007688/1).
Computing facilities were provided by the Scientific Computing Research Technology Platform of the University of Warwick through the use of the High Performance Computing (HPC) cluster Avon, and the Sulis Tier 2 platforms at HPC Midlands+ funded by the Engineering and Physical Sciences Research Council (EPSRC), grant number EP/T022108/1. Computational support was also obtained from the UK national high performance computing service, ARCHER2, for which access was obtained via the UKCP consortium and funded by EPSRC grant ref EP/X035891/1. We thank Diamond Light Source for access to beamline I05 under proposals SI34479, SI33317, and SI35210.

\section*{References}

\begin{flushleft}
\bibliographystyle{iopart-num}
\bibliography{references,theory_refs}
\end{flushleft}

\end{document}


\newcommand{\WSe}{WSe\textsubscript{2}~}

\title[SI:WSe$_2$ on Au]{Supplementary data for: Electronic structure of the interface between Au and WSe$_2$}

\author{Laxman Nagireddy, Samuel J. Magorrian, Matthew D. Watson, Yogal Prasad Ghimirey, Cephise Cacho, Neil R. Wilson and Nicholas D.M. Hine}
\ead{N.D.M.Hine@warwick.ac.uk}\ead{Neil.Wilson@warwick.ac.uk}

\vspace{2pc}
\noindent{{\it \today}}



\section{Core level spectroscopy} 

The peak positions found from analysis of the core level spectra of \WSe on Si-TSG (Figure 6 of the main text) are given in table \ref{tab:core level Si tsg}. These values are given as binding energies, where the binding energy $\text{BE}=-\left(\text{E}-\text{E}_\text{F}\right)$.

\renewcommand{\arraystretch}{2}
\begin{table}[ht]
    \centering
    \scriptsize
\begin{tabular}{|c|c|c|c|c|c|}
\hline
Scan region & 1L (eV) & 2L (eV) & 3L (eV)  & Au (eV)  \\ 
\hline
W-4f$_{5/2}$ & $34.56 \pm 0.05$ & $34.85 \pm 0.01$ / $34.64 \pm 0.02$ & $35.03 \pm 0.02$ / $34.70 \pm 0.05$ & \\ 
\cline{1-2} 
W-4f$_{7/2}$ & $32.43 \pm 0.01$ & $32.72 \pm 0.02$ / $32.51 \pm 0.01$ & $32.91 \pm 0.01$ / $32.60 \pm 0.05$ & \\ 
\hline
$\Delta$SOC W & $2.18 \pm 0.05$ & $2.13 \pm 0.02$ & $2.12 \pm 0.02$ / $2.10 \pm 0.07$ &\\ 
\hline
Se-3d$_{3/2}$ & $55.56 \pm 0.02$ & $55.85 \pm 0.05$ / $55.62 \pm 0.02$ & $55.98 \pm 0.02$ / $55.66 \pm 0.02$ & \\ 
\hline
Se-3d$_{5/2}$ & $54.71 \pm 0.05$ & $55.05 \pm 0.07$ / $54.79 \pm 0.02$ & $55.13 \pm 0.01$ / $54.81 \pm 0.03$ & \\ 
\hline
$\Delta$SOC Se & $0.85 \pm 0.05$ & $0.80 \pm 0.08$ / $0.79 \pm 0.02$ & $0.83 \pm 0.02$ / $0.79 \pm 0.03$ & \\ 
\hline
Au-4f$_{5/2}$ & $87.63 \pm 0.02$ & $87.63 \pm 0.04$ & & $87.56 \pm 0.01$   \\ 
\hline
Au-4f$_{7/2}$ & $83.94 \pm 0.01$ & $83.93 \pm 0.02$ & & $83.90 \pm 0.05$   \\ 
\hline
$\Delta$SOC Au & $3.69 \pm 0.02$ & $3.70 \pm 0.04$ & & $3.66 \pm 0.0$  \\ 
\hline
Si 2p & $102.56 \pm 0.05$ & & & $102.12 \pm 0.08$   \\ 
\hline
\end{tabular}

    \caption{Fitting parameters from core level spectra on Si-TSG (Figure 6 of the main text).}
    \label{tab:core level Si tsg}
\end{table}

\begin{figure}[t]   
\centering
        \includegraphics[width=0.8\linewidth]{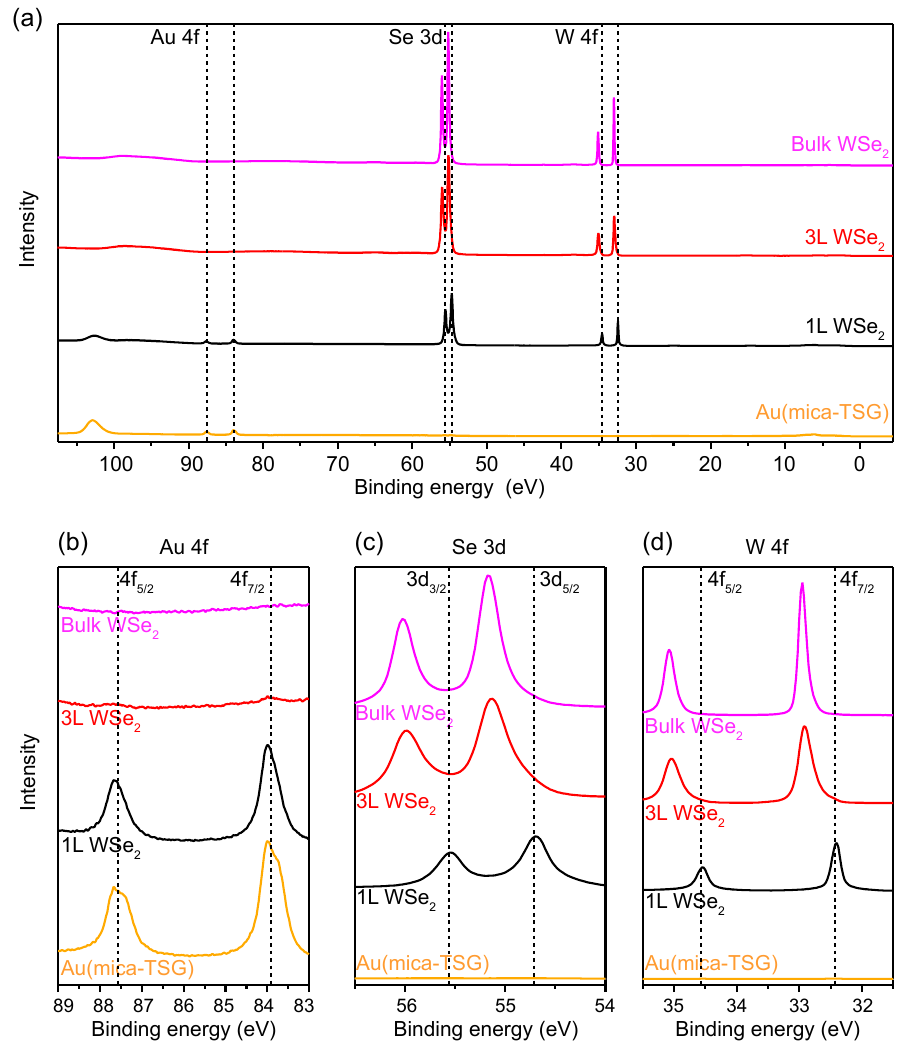}
    \caption{\textbf{Core level spectra from WSe$_2$ on mica-TSG}. Survey scan, (a), and region scans of Au 4\textit{f}, Se 3\texit{d}, and W 4\textit{f}, acquired at a photon energy of 140 eV.}
    \label{fig:SImicacorelevel}
\end{figure}

Core level spectra from WSe$_2$ on mica-TSG are shown in Figure \ref{fig:SImicacorelevel}. The corresponding peak positions are given in table \ref{tab:MicaTSG Corelevel spectra}. The same core conclusions can be drawn from these data as from the specrta from WSe$_2$ on Si-TSG shown in the main text: the interfacial layer of WSe$_2$ is in a different chemical environment to the subsequent ones, and there is only one chemical environment for the Se in the interfacial layer with no discernible difference between upper and lower Se atoms.

\renewcommand{\arraystretch}{2}
\begin{table}[]
    \centering
    \scriptsize
    \begin{tabular}{|c|c|c|c|c|c|}
\hline
Scan region & 1L (eV) & 3L (eV) & Bulk (eV) & Au (eV) \\ 
\hline
W-4f$_{5/2}$ & $34.55 \pm 0.01$ & $35.03 \pm 0.01$ & $35.07 \pm 0.01$ & \\ 
\hline
W-4f$_{7/2}$ & $32.42 \pm 0.01$ & $32.89 \pm 0.02$ & $32.94 \pm 0.05$ & \\ 
\hline
$\Delta$SOC W & $2.13 \pm 0.014$ & $2.14 \pm 0.02$ & $2.14 \pm 0.05$ & \\ 
\hline
Se-3d$_{3/2}$ & $55.54 \pm 0.01$ & $56.0 \pm 0.02$ / 55.84 $\pm$ 0.05 & $55.15 \pm 0.02$ & \\ 
\hline
Se-3d$_{5/2}$ & $54.70 \pm 0.02$ & $55.14 \pm 0.02$ /54.94 $\pm$ 0.03 & $56.01 \pm 0.02$ & \\ 
\hline
$\Delta$SOC Se & $0.84 \pm 0.02$ & $0.86 \pm 0.03$ / $0.9 \pm 0.06 $ & $0.86 \pm 0.02$ & \\ 
\hline
Au-4f$_{5/2}$ & $87.60 \pm 0.01$ & & &  $87.56 \pm 0.01$ \\ 
\hline
Au-4f$_{7/2}$ & $83.95 \pm 0.01$ & & &  $83.90 \pm 0.05$ \\ 
\hline
$\Delta$SOC Au & $3.65 \pm 0.02$ & & &  $3.66 \pm 0.05$ \\ 
\hline
Si 2p & $102.54 \pm 0.05$ & & & $102.75 \pm 0.10$ \\ 
\hline
\end{tabular}
    \caption{Fitting parameters from core level spectra on mica-TSG.}
    \label{tab:MicaTSG Corelevel spectra}
\end{table}

\clearpage

\section*{Density of States as a function of layer number}

To determine thickness required for reasonable convergence to the bulk-like limit, calculations were performed on the isolated slab model of the Au(111) surface, for models comprising 3, 6 and 9 layers. It was determined that 6 layers was sufficient to converge peak positions and general shape sufficiently close to the bulk-like limit, as seen in Supplementary Figure \ref{fig:AuSlabDOS}.

\begin{figure}[t]   
\centering
        \includegraphics[width=0.8\linewidth]{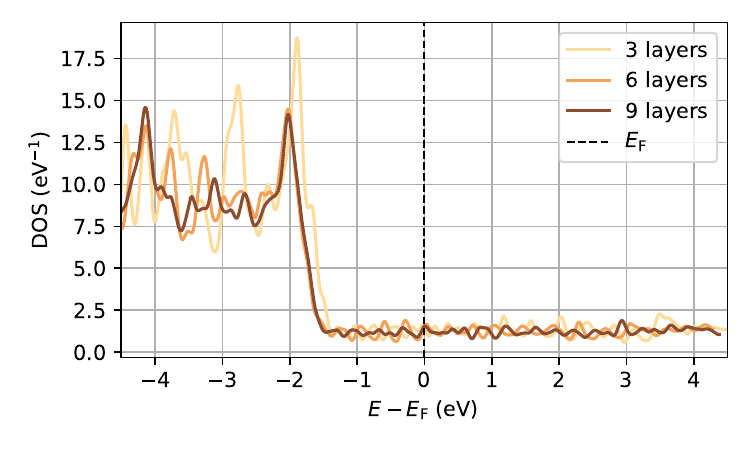}
    \caption{Density of States plot for Au(111) slab models of 3, 6 and 9 layers.}
    \label{fig:AuSlabDOS}
\end{figure}
\clearpage


